\documentclass[journal]{IEEEtran} % use the `journal` option for ITherm conference style
\IEEEoverridecommandlockouts
\usepackage{cite}
\usepackage{amsmath,amssymb,amsfonts}
\usepackage{amsthm}
\usepackage{algorithmic}
\usepackage{subcaption}
\usepackage{multicol}
\usepackage{graphicx}
\usepackage{textcomp}
\usepackage{bbm}
\usepackage[hidelinks]{hyperref}
\usepackage{xcolor}

\usepackage[a4paper,bindingoffset=0.2in,
            left=0.65in,right=0.65in,top=0.75in,bottom=1.1in,
            footskip=.25in]{geometry}

\newtheorem{definition}{Definition}
\newtheorem{lemma}{Lemma}
\def\BibTeX{{\rm B\kern-.05em{\sc i\kern-.025em b}\kern-.08em
    T\kern-.1667em\lower.7ex\hbox{E}\kern-.125emX}}
\begin{document}

\title{An Optimization Framework For Anomaly Detection Scores Refinement With Side Information
% delete or comment-out the following line before submission
}

\author{%%%% author names
    \IEEEauthorblockN{\textnormal{Ali Maatouk}$^*$}% first author
    , \IEEEauthorblockN{\textnormal{Fadhel Ayed}$^*$}% delete this line if not needed
    , \IEEEauthorblockN{\textnormal{Wenjie Li}$^*$}% delete this line if not needed
    , \IEEEauthorblockN{\textnormal{Yu Wang}$^\mathsection$}% delete this line if not needed
    , \IEEEauthorblockN{\textnormal{Hong Zhu}$^\mathsection$}% delete this line if not needed
    , \IEEEauthorblockN{\textnormal{Jiantao Ye}$^\dagger$}% delete this line if not needed
    % duplicate the line above as many times as needed to list all authors
    \\%%%% author affiliations
    \IEEEauthorblockA{$^*$Paris Research Center, Huawei Technologies, Boulogne-Billancourt, France}\\% first affiliation
    \IEEEauthorblockA{$^\mathsection$China United Network Communications Group Co., Ltd, Nanjing, China}\\% delete this line if not needed
    % duplicate the line above as many times as needed to list all affiliations
    %%%% corresponding author contact details
    \IEEEauthorblockA{$^\dagger$Songshan Lake Research Center, Huawei Technologies, Dongguan, China}
}

\maketitle

\begin{abstract} This paper considers an anomaly detection problem in which a detection algorithm assigns anomaly scores to multi-dimensional data points, such as cellular networks' Key Performance Indicators (KPIs). We propose an optimization framework to refine these anomaly scores by leveraging side information in the form of a causality graph between the various features of the data points. The refinement block builds on causality theory and a proposed notion of confidence scores. After motivating our framework, smoothness properties are proved for the ensuing mathematical expressions. Next, equipped with these results, a gradient descent algorithm is proposed, and a proof of its convergence to a stationary point is provided. Our results hold (i) for any causal anomaly detection algorithm and (ii) for any side information in the form of a directed acyclic graph. Numerical results are provided to illustrate the advantage of our proposed framework in dealing with False Positives (FPs) and False Negatives (FNs). Additionally, the effect of the graph's structure on the expected performance advantage and the various trade-offs that take place are analyzed.
\end{abstract}

%\begin{IEEEkeywords}
 %   component, formatting, style, styling, insert
%\end{IEEEkeywords}

\section{Introduction}
Anomaly detection refers to the problem of finding patterns in data that do not
conform to a notion of normal behavior. These detected patterns are typically referred to as anomalies, outliers, or contaminants depending on the application at hand. The importance of such frameworks is that anomalies in data
translate to critical information about the status of the system at hand in a wide variety of application domains. For example, in a cellular network, anomalies in the Key Performance Indicators (KPIs) such as Downlik/Uplink (DL/UL) throughput can be a sign of a hardware/software malfunction or malicious activity \cite{https://doi.org/10.1049/iet-com.2019.0765}. Traditionally, the identification of such anomalies was carried on by network engineers. However, such a process is both costly and time-consuming for network operators. Additionally, 5G networks are expected to provide 99.999\%, or “five nines,” of data availability annually \cite{9700502}, set to improve to a seven-nines standard in 6G \cite{9349624}. To have a grasp on how stringent this requirement is, 5G networks are expected to have just six minutes of unscheduled downtime per year.

%Given these strict requirements and the foreseen complexity in operating and managing 5G and beyond networks, a trend toward closed-loop automation of network and service management operations can be witnessed. Particularly, a Zero-touch network and Service Management (ZSM) framework is envisaged as a
%next-generation management system that aims to
%have all operational processes and tasks executed
%automatically \cite{8994961}. Note that a major block of the ZSM is the automatic detection of anomalies. \color{red}To achieve this end, a flurry of unsupervised anomaly detection algorithms are available, building on the most state-of-the-art advances in AI to monitor the health of complex systems \cite{ren2019time,carmona2021neural,ayed2020anomaly,gao2020robusttad}. 

With these strict requirements and anticipated complexity in managing 5G and beyond networks, a shift towards closed-loop automation of network and service management operations is underway. The Zero-touch network and Service Management (ZSM) framework is envisioned as the next-generation management system that aims to automate all operational processes and tasks \cite{8994961}. The automatic detection of anomalies is a major component of the ZSM, which can be achieved using a flurry of unsupervised anomaly detection algorithms based on state-of-the-art advances in AI \cite{ren2019time,carmona2021neural,ayed2020anomaly,gao2020robusttad}. For example, \cite{ren2019time} proposed a spectral residual method based on Fast Fourier Transform, which achieved state-of-the-art performance on Microsoft production data. Additionally, studies such as \cite{carmona2021neural,ayed2020anomaly,gao2020robusttad} have proposed using convolutional and recurrent neural networks to continuously monitor time-series data and alert for potential anomalies. In addition to the methods mentioned above, there are other basic anomaly detectors in the literature that relies on statistical measures to identify anomalies in the data and are relatively simple to implement \cite{10.1145/1541880.1541882}.

% The automatic detection of cellular networks anomalies relies on network measurements and parameters, which are stored and sometimes exchanged between the User equipment (UE) and the base stations. 

Regardless of the anomaly detection method used, there are common challenges in accurately detecting anomalies in cellular networks. The first challenge is defining the normal behavior of the KPIs and identifying any observations that deviate from this behavior as anomalous. Moreover, the boundary between normal and anomalous behavior is often not well-defined, making it challenging to identify and classify anomalies accurately. Another significant challenge is the availability of labeled data for validating the anomaly detection models. In many cases, labeled data is scarce, making it difficult to build accurate models that can identify even rare and subtle anomalies \cite{10.1145/1541880.1541882}. Irrespective of the selected anomaly detector, these challenges often lead to False Positives (FPs) and False Negatives (FNs), which can significantly impact the performance of the anomaly detection framework. Therefore, improving the False Positive Rate (FPR) and False Negative Rate (FNR) is a necessary step to ensure the efficient operation of the anomaly detection block. To this end, a crucial question arises: Given a chosen anomaly detector, can we leverage experts' knowledge about the relationship between the various KPIs to enhance the FPR and FNR in the context of cellular networks? This is the problem that we tackle in our paper.
\color{black}
Specifically, our contributions can be summarized below:
\begin{itemize}
\item We introduce a mathematical framework for refining the anomaly score at the output of any generic causal anomaly detector by leveraging available side information. Particularly, we consider that the side information consists of a directed acyclic graph that represents the various causality relationships between the KPIs. Additionally, we introduce the notion of KPI confidence score and provide the motivation behind each component of our framework.
\item Next, we analyze the properties of the ensuing mathematical expressions. Particularly, we show that smoothness bounds are verified and, hence, we propose a gradient descent algorithm to solve our non-convex optimization problem. Afterward, we theoretically prove the convergence of our algorithm to a stationary point.
\item Lastly, we implement our proposed optimization framework when the directed acyclic graph belongs to the family of polytrees. Particularly, we demonstrate that adding our refinement block significantly enhances the performance of the anomaly detection framework in this case. The degree of improvement is shown to be dependent on the structure of the graph, and we present simulations that illustrate the tradeoffs at play. To emphasize the practical benefits of our approach, we implement our framework using a directed acyclic graph developed by our team that captures all the intrinsic causality relationships between a cellular network' KPIs. We show that this practical implementation yields similar performance advantages. 
\end{itemize}

%Apparently, automatic anomaly detection is critical for effective operation and cellular network management. This is generally achieved based on network measurements and parameters. In cellular networks (e.g. LTE/LTE-A), there are lots of network parameters and measurements, which are consecutively exchanged, collected, and reported at/from the nodes and user equipment (UE) in the core networks and the radio access network, such as call detail record (CDR), location information, UE movement behaviour, reference signal received power, radio link failure report, and so on. We can detect network anomalies through the analysis of the network parameters and measurements, and thus, the network behaviour can be monitored to avoid possible threats, failure, and faults. The general system architecture of anomaly detection is shown in Fig. 1. Overall, anomaly detection in cellular networks has three major advantages. First, it facilitates the network operator to perform effective management of the cellular networks. Second, it enables the optimization and enhancement of user QoE by exploring relevant historical information. Third, network-wide insights can be obtained to facilitate efficient network planning and deployment. Owing to the above benefits, it is very crucial to timely deal with the potential network anomalies.

The rest of the paper is organized as follows: Section \ref{systmodel}
is dedicated to the system model. The proposed optimization framework is presented in Section \ref{optimizframework}. Section \ref{propsolution} provides a description of our proposed solution to the optimization problem. Numerical results that corroborate our theoretical findings are laid out in Section \ref{numericalstuff}, while the paper is concluded in Section \ref{concsection}.
\section{System Model}
\label{systmodel}
\begin{figure}[!t]
    \centerline{\includegraphics{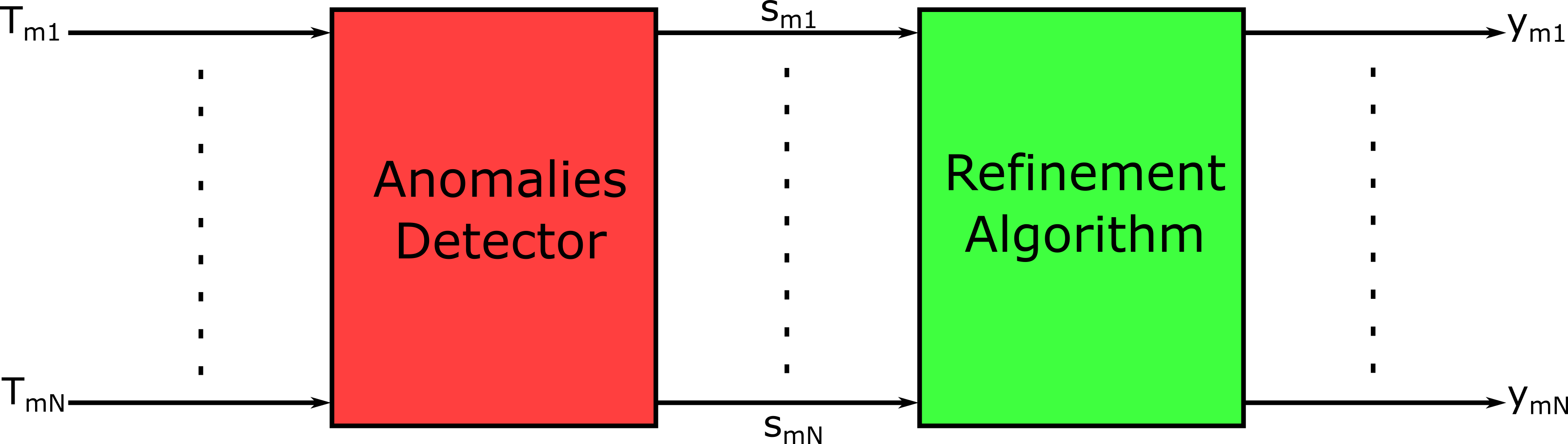}}
        \caption{Illustration of the proposed framework.}
    \label{figillustration}
    \vspace{-10pt}
\end{figure}
A time series (or data series in general) is an ordered set $\mathcal{T}=\{\boldsymbol{T}_1,\boldsymbol{T}_2,\ldots,\boldsymbol{T}_M\}$ of $M$ real-valued, potentially multi-dimensional data points $\boldsymbol{T}_m\in\mathbb{R}^N$. In the cellular networks' framework, these time series can represent the various KPIs monitored at a small time granularity by the cellular sites. These KPIs may include, for example, the UL/DL throughput, the call drop rate, the handover success rate, etc. Such KPIs provide key information about the status of the network and are considered fundamental indicators of any anomalous behavior taking place, thus allowing the network operator to initiate troubleshooting efforts when necessary. To detect these anomalous behaviors automatically, a time series anomaly detection framework that marks anomalies in $\mathcal{T}$ is required. With this in mind, let us define the score set $\mathcal{S}=\{\boldsymbol{s}_1,\boldsymbol{s}_2,\ldots,\boldsymbol{s}_M\}$ with $\boldsymbol{s}_m\in[0,1]^N$ being the result of a time series anomalies detection algorithm that assigns for each data point $T_{mi}$ for $i=1,\ldots,N$, an anomaly score $s_{mi}\in[0,1]$. The score can be interpreted as the probability of an anomalous behavior occurring in the corresponding time series value. Note that for any two scores $s_{mi}$ and $s_{mj}$, it must be true that if $s_{mi}>s_{mj}$, then $T_{mi}$ is more anomalous than $T_{mj}$ in their respective contexts. Now, generally, the dimension of the time series $N$ is large and the behavior of the KPIs can be quite heterogeneous. Additionally, missing values can occur during data acquisition for a subset of the time series $\mathcal{T}$. With this in mind, and given the typical scarcity of labeled data in anomaly detection frameworks, multi-dimensional time series anomaly detection algorithms can perform poorly \cite{Thudumu2020}. Therefore, we consider in the following that the score $s_{mi}$ is calculated as follows
\begin{equation}
s_{mi}=f_i(T_{1i},\ldots,T_{mi}), \quad \textnormal{for } i=1,\ldots,N,
\end{equation}
where $f_i$ represents the set of rules and procedures of the anomaly detector $i$. Note that given the time-causality principle, the score $s_{mi}$ can only depend on the current value $T_{mi}$ and its history. 

The goal of our work is to refine the score set $\mathcal{S}$ by outputting a set of final scores $\mathcal{Y}=\{\boldsymbol{y}_1,\boldsymbol{y}_2,\ldots,\boldsymbol{y}_M\}$, with $\boldsymbol{y}_m\in[0,1]^N$ as shown in Fig. \ref{figillustration}. To do such a refinement, we leverage available experts' side knowledge consisting of causality relationships between the various KPIs. With this in mind, the challenge consists of establishing a mathematical framework for such refinement and elaborating the necessary optimization techniques. Note that our score refinement framework is general as it imposes no restrictions on the anomalies detection algorithm, hence making it applicable to various types of anomaly detectors. 
%As an example of such relationships, if the KPI representing the Channel Quality Indicator (CQI) falls drastically, that would also lead to anomalous behavior in the KPI representing the throughput performance. Accordingly, a causality relationship between these two KPIs exists. 
\section{Optimization Framework}
\label{optimizframework}
To begin our analysis, we represent the expert knowledge regarding the network KPIs' behavior as a directed acyclic graph, denoted by $G=(V,E)$. Here, the set $V$ consists of the network's key KPIs, where $|V|=N$. The set of edges, $E$, indicates the causal relationships between the various KPIs. 
 In particular, a directed edge between two nodes $i_1$ and $i_2$ means that an anomalous behavior in KPI $i_1$ is typically caused by an anomaly in KPI $i_2$. For instance, if the KPI representing Channel Quality Indicator (CQI) drops dramatically, it can lead to anomalous behavior in the KPI representing throughput performance. Therefore, a directed edge originating from the node representing throughput to the one representing CQI should exist in $G$. Our team has developed an expert graph with over 50 nodes and 100+ relationships between the different cellular network KPIs, which we will leverage in our implementations in Section \ref{expertsgraphimplementation}. Next, to pursue our analysis, we define the set of neighbors of each node $i\in V$ as
\begin{equation}
    \mathcal{N}_i=\{j:(i,j)\in E\}.
\end{equation}
%is reported in Fig. \ref{figexperts}, where 
%the leaf nodes of the graph (i.e., $\mathcal{L}=\{i: \mathcal{N}_i=\emptyset\}$) are colored in green. 
We also introduce a confidence score $\boldsymbol{\alpha}_m\in[0,1]^N$ for $m=1,\ldots,M$ that represents our degree of confidence in the score set $\mathcal{S}$. The grounds for this confidence score are threefold:
\begin{itemize}
\item A set of KPIs $\Lambda\subset V$ are typically considered ``Key" KPIs in the sense that their scores are robust. For such KPIs, two elements are true: robustness in measurements, and simplicity of the anomaly detection rules. Accordingly, the anomaly scores at the output of their respective detectors can be fully trusted (i.e., $\alpha_{mi}=1$ for $i\in\Lambda$ and $m=1,\ldots,M$). 
    \item If the time series data are missing for a particular set of KPIs $\Delta_m$ at a time slot $m$, then we set $\alpha_{mi}$ to $0$ for $i\in\Delta_m$. By doing so, the anomaly scores of these detectors are not trusted. Correspondingly, the emphasis is put on the experts' graph to rectify the scores $s_{mi}$ for $i\in\Delta_m$ if needed. 
    \item Between the above two extremes, the confidence score is a variable that needs to be optimized. Particularly, it allows us to put more emphasis on the experts' knowledge rather than the score set $\mathcal{S}$ and vice-versa as necessary. 
    %Accordingly, we define $\tilde{\boldsymbol{\alpha}}_m\in[0,1]^{N-|\Lambda|-|\Delta_m|}$ that consists of all the nodes that sit between the two extremes mentioned above.
\end{itemize}

%\begin{figure}[!t]
%    \centerline{\includegraphics[width=.55\linewidth]{subgraph.png}}
%%        \caption{Illustration of the experts' graph.}
%    \label{figexperts}
%%    \vspace{-10pt}
%\end{figure}

Lastly, before formulating our optimization framework, we provide several key requirements that the refinement block reported in Fig. \ref{figillustration} should obey at each time epoch $m$: 
\begin{itemize}
    \item The output $\boldsymbol{y}_m$ should be close to $\boldsymbol{s}_m$. Given that the goal is to refine the score vector $\boldsymbol{s}_m$, the optimization framework should not completely disregard $\boldsymbol{s}_m$ but rather keep $\boldsymbol{y}_m$ close to it. 
    \item The confidence score at each time epoch $m$ should not be too small (i.e., $\alpha_{mi}\geq \overline{\alpha}$ for $i\in V\setminus\Delta_m$, where $\overline{\alpha}>0$). This constraint also prevents the optimization algorithm from fully disregarding the score vector $\boldsymbol{s}_m$.
    \item For each node $i\in V: \mathcal{N}_i\neq\emptyset$, the output vector $\boldsymbol{y}_m$ should verify the following constraint 
    \begin{equation}
        y_{mi}\leq \max_{j\in\mathcal{N}_i} y_{mj}. 
        \label{constraintneighborhood}
    \end{equation}
    The above constraint means that if anomalous behavior is detected in KPI $i$, then at least one of its neighbors should be found as anomalous.
\end{itemize}
With the above requirements in mind, we can formulate our optimization problem at each time epoch $m$ as follows
\begin{equation}
\setlength{\belowdisplayskip}{0pt} \setlength{\belowdisplayshortskip}{0pt}
\setlength{\abovedisplayskip}{0pt} \setlength{\abovedisplayshortskip}{0pt} 
\begin{aligned}
&\underset{\boldsymbol{y}_m,\boldsymbol{\alpha}_m}{\text{minimize}}
&& f(\boldsymbol{y}_m, \boldsymbol{\alpha}_m)=\frac{\sum_{i=1}^{N}\alpha_{mi}(s_{mi}-y_{mi})^2}{\sum_{i=1}^{N}\alpha_{mi}}\\
& \text{s.t.}
&&  y_{mi}\leq \max_{j\in\mathcal{N}_i} y_{mj}, \quad \text{for}\:\: i\in V: \mathcal{N}_i\neq\emptyset,\\
& && y_{mi}\in[0,1], \quad \text{for}\:\: i\in V,\\ 
& && \alpha_{mi}\in[\overline{\alpha},1],\quad \text{for}\:\: i\in V\setminus\Delta_m\cup\Lambda,\\ 
& && \alpha_{mi}=1, \quad \text{for}\:\: i\in \Lambda,\\ 
& && \alpha_{mi}=0, \quad \text{for}\:\: i\in \Delta_m.
\end{aligned}
\label{constrainedobjective}
\end{equation}
In essence, $f(\boldsymbol{y}_m, \boldsymbol{\alpha}_m)$ ensures that the output score $\boldsymbol{y}_m$ is not far from $\boldsymbol{s}_m$. Additionally, the normalization by $\sum_{i=1}^{N}\alpha_{mi}$ ensures that the optimization of $f(\boldsymbol{y}_m, \boldsymbol{\alpha}_m)$ does not force $\alpha_{mi}$ to be equal to $\overline{\alpha}$ for $i\in V\setminus\Delta_m\cup\Lambda$. On the other hand, the constraints ensure that the requirements that we have previously set are met. Note that, since the values of $\alpha_{mi}$ are fixed for $i\in\Lambda\cup\Delta_m$, we remove them from our optimization variables. Accordingly, in the sequel, the optimization variable $\boldsymbol{\alpha}_m$ will only be defined for $i\in V\setminus\Delta_m\cup\Lambda$, while $\boldsymbol{\alpha}_m$ for $i\in \Delta_m\cup\Lambda$ will be fixed. 

Next, we aim to transform our constrained optimization problem into an unconstrained one. To do so, we first rewrite the variables $y_{mi}$ and $\alpha_{mi}$ as follows 
\begin{equation}
y_{mi}=\sigma(z_{mi})=\frac{1}{1+\exp(-z_{mi})},
\end{equation}
\begin{align}
\alpha_{mi}=\overline{\alpha}+(1-\overline{\alpha})\sigma(\varepsilon_{mi})=&\overline{\alpha}+(1-\overline{\alpha})\frac{1}{1+\exp(-\varepsilon_{mi})},
\end{align}
where $\sigma(\cdot)$ denotes the sigmoid function and $(z_{mi},\varepsilon_{mi})$ are the new optimization variables. By doing so, we ensure that the constraints on the space of $\boldsymbol{y}_m$ and $\boldsymbol{\alpha}_m$ are met. It is worth noting that this approach is a common practice in machine learning applications, and is widely used in various settings (e.g., logistic regression, output layer of a neural network, etc. \cite{shalev-shwartz_ben-david_2014}). As a last step, we aim to provide a smoothed version of the constraint reported in eq. (\ref{constraintneighborhood}). To that end, we introduce the following approximation
\begin{equation}
\max_{j\in\mathcal{N}_i} y_{mj}\approx\frac{\sum_{j\in\mathcal{N}_i}y_{mj}\exp(cy_{mj})}{\sum_{j\in\mathcal{N}_i}\exp(cy_{mj})}.
\label{softmax}
\end{equation}
The accuracy of such approximation increases with $c$. However, the smoothness property of the function goes worse when increasing $c$, making the function harder to be optimized. In the following, we will consider that $c=10$, which achieves an adequate trade off between smoothness and accuracy. With this in mind, our final optimization problem can be summarized below
\begin{equation}
\setlength{\belowdisplayskip}{0pt} \setlength{\belowdisplayshortskip}{0pt}
\setlength{\abovedisplayskip}{0pt} \setlength{\abovedisplayshortskip}{0pt} 
\begin{aligned}
&\underset{\boldsymbol{z}_m, \boldsymbol{\varepsilon}_m}{\text{minimize}}
&& \hspace{-140pt}\overline{f}(\boldsymbol{z}_m, \boldsymbol{\varepsilon}_m)=\\ & \frac{\sum_{i=1}^{N}\overline{\alpha}+(1-\overline{\alpha})\sigma(\varepsilon_{mi})(s_{mi}-\sigma(z_{mi}))^2}{\sum_{i=1}^{N}\overline{\alpha}+(1-\overline{\alpha})\sigma(\varepsilon_{mi})}&& \\ & +\mu\sum_{i:\mathcal{N}_i\neq\emptyset}&& \hspace{-140pt}\big([\sigma(z_{mi})-\frac{\sum_{j\in\mathcal{N}_i}\sigma(z_{mj})\exp(c\sigma(z_{mj}))}{\sum_{j\in\mathcal{N}_i}\exp(c\sigma(z_{mj}))}]_{+}\big)^3.
\end{aligned}
\label{finalunconstrainedobjective}
\end{equation} 
Note that the term $[\sigma(z_{mi})-\frac{\sum_{j\in\mathcal{N}_i}\sigma(z_{mj})\exp(c\sigma(z_{mj}))}{\sum_{j\in\mathcal{N}_i}\exp(c\sigma(z_{mj}))}]_{+}$ was raised to the cube to ensure the twice differentiability of $\overline{f}$ and to enhance the smoothness  properties of $\overline{f}$, as will be seen later in the proof of Lemma \ref{lemmabeta}. On the other hand, the multiplier $\mu$ is a fixed value that can be interpreted as a cost for the violation of the corresponding constraints. Particularly, we increase $\mu>0$ the more trustworthy the experts' graph is. In the sequel, we will assign a large fixed value to $\mu$, hence completely trusting our experts' graph. With the optimization problem being formulated, we can now tackle it and provide a time-efficient solution to it. 
\section{Proposed Solution}
\label{propsolution}
\begin{table*}[ht]
\begin{equation*}
    \frac{\partial \overline{f}}{\partial y_{mk}}= \frac{2\alpha_{mk}(y_{mk}-s_{mk})}{\sum_{i=1}^{N}\alpha_{mi}}+3\mu\sum_{i:\mathcal{N}_i\neq\emptyset}[y_{mi}-\frac{\sum_{j\in\mathcal{N}_i}y_{mj}\exp(cy_{mj})}{\sum_{j\in\mathcal{N}_i}\exp(cy_{mj})}]^2_{+}\big[\mathbbm{1}\{k=i\}-\mathbbm{1}\{k\in\mathcal{N}_i\}\times
\end{equation*}
\vspace{-25pt}
\begin{multicols}{2}
\begin{equation}
    \frac{\exp(cy_{mk})\big(1+cy_{mk}-c\frac{\sum_{j\in\mathcal{N}_i}y_{mj}\exp(cy_{mj})}{\sum_{j\in\mathcal{N}_i}\exp(cy_{mj})}\big)}{\sum_{j\in\mathcal{N}_i}\exp(cy_{mj})}\big].
    \label{firsteq}
\end{equation}\break
\begin{equation}
     \frac{\partial \overline{f}}{\partial \alpha_{mk}}=\frac{\sum_{i=1}^{N}\alpha_{mi}\big((s_{mk}-y_{mk})^2-(s_{mi}-y_{mi})^2\big)}{\sum_{i=1}^{N}\alpha_{mi}}.
    \label{secondeq}
\end{equation}
\end{multicols}

%\begin{align}
  %  \frac{\partial \overline{f}}{\partial y_{mk}}= \frac{2\alpha_{mk}(y_{mk}-s_{mk})}%{\sum_{i=1}^{N}\alpha_{mi}}&+3\mu\sum_{i:\mathcal{N}_i\neq\emptyset}[y_{mi}-%\frac{\sum_{j\in\mathcal{N}_i}y_{mj}\exp(cy_{mj})}%{\sum_{j\in\mathcal{N}_i}\exp(cy_{mj})}]^2_{+}\big[\mathbbm{1}\{k=i\}-\mathbbm{1}\{k\in\mathcal{N}_i\}\times\nonumber\\&
 %   \frac{\exp(cy_{mk})\big(1+cy_{mk}-c\frac{\sum_{j\in\mathcal{N}_i}y_{mj}\exp(cy_{mj})}%{\sum_{j\in\mathcal{N}_i}\exp(cy_{mj})}\big)}{\sum_{j\in\mathcal{N}_i}\exp(cy_{mj})}\big].
%    \label{firsteq}
%\end{align}

\vspace{-25pt}
\begin{multicols}{2}
\begin{equation}
    \frac{\partial y_{mk}}{\partial z_{mk}}=\frac{\exp(-z_{mk})}{\big(1+\exp(-z_{mk})\big)^2}.
    \label{thirdeq}
\end{equation}\break
\begin{equation}
    \frac{\partial \alpha_{mk}}{\partial \varepsilon_{mk}}=(1-\overline{\alpha})\frac{\exp(-\varepsilon_{mk})}{\big(1+\exp(-\varepsilon_{mk})\big)^2}.
    \label{fourtheq}
\end{equation}
\end{multicols}
\caption{Details of the gradient of $\overline{f}(\boldsymbol{z}_m, \boldsymbol{\varepsilon}_m)$.}
\label{gradientable}
 \vspace{-15pt}
\end{table*}

In this section, we introduce our proposed solution to the optimization problem reported in eq. (\ref{finalunconstrainedobjective}). First, we note that the optimization problem reported in eq. (\ref{finalunconstrainedobjective}) is non-convex. Additionally, we recall that the refinement algorithm needs to run at each time-epoch $m$. Given that the time-granularity of the KPIs can be small, the low-complexity of the refinement algorithm becomes a critical requirement. With this in mind, we first proceed with finding the gradient of the objective function $\overline{f}(\boldsymbol{z}_m, \boldsymbol{\varepsilon}_m)$. With the derivatives chain rule in mind, the gradient can be concluded from the equations reported in Table \ref{gradientable}. Next, we investigate the smoothness property of our function $\overline{f}(\boldsymbol{z}_m, \boldsymbol{\varepsilon}_m)$. Before doing so, we note that the case where $\Delta_m=V$ (i.e., all scores are not trustworthy) is not of interest. Accordingly, we focus below only on the case where $\Delta_m\neq V$.
\begin{lemma}
Let the smoothness factor $\beta$ denote the maximum eigenvalue, in magnitude, of the Hessian matrix $\boldsymbol{H}_{\overline{f}}$ over $\mathbb{R}^{2N}$ (domain of $(\boldsymbol{z}_m, \boldsymbol{\varepsilon}_m)$). Given the structure of $\overline{f}$ reported in eq. (\ref{finalunconstrainedobjective}), and if $\Delta_m\neq V$, then there exists a constant $L\in[0,\infty]$ such that $\beta\leq L$.
\label{lemmabeta}
\end{lemma}
\begin{IEEEproof}
The proof can be found in Appendix \ref{prooflemmabeta}.
\end{IEEEproof}
Given the above results, and the fact that $\overline{f}(\cdot,\cdot)$ is twice differentiable with continuous derivatives, we can conclude from the descent lemma that, for a sufficiently small step size $\gamma\leq \frac{1}{L}$, the gradient descent algorithm converges to a stationary point with a convergence rate of $\mathcal{O}(1/t)$ \cite{Cunha13}. To that end, we present below our proposed iterative optimization algorithm: 
\begin{equation}
z_{mi,t+1}=z_{mi,t}-\gamma\frac{\partial \overline{f}}{\partial z_{mi,t}},
\end{equation}
\begin{equation}
\varepsilon_{mi,t+1}=z_{mi,t}-\gamma\frac{\partial \overline{f}}{\partial \varepsilon_{mi,t}},
\end{equation}
where $z_{mi,0}$ and $\varepsilon_{mi,0}$ are initialized from a standard Gaussian distribution $\mathcal{N}(0,1)$. In the next section, we will implement our algorithm and showcase the performance advantage it provides in various settings. 
\section{Numerical Implementations}
\label{numericalstuff}
\subsection{Performance Comparison}
\label{performancecomparisonsection}
In this section, we will focus on implementing numerically our proposed refinement block and highlighting its performance advantage. To that end, let us first delve into the details of the considered experts' graph $G=(V,E)$. Particularly, we will consider in this section that the graph $G$ belongs to the perfectly balanced polytrees family defined below.
\begin{definition}[Perfectly Balanced $(r,h)$-Polytree] A directed acyclic graph $G=(V,E)$ is said to be a perfectly balanced $(r,h)$-polytree if its underlying undirected graph is a tree in which every internal node has exactly $r$ child nodes and all the leaf nodes are at the same height level $h$.
\end{definition}
Following the above definition, we can conclude that the number of nodes $N$ of an $(r,h)$-Polytree is
\begin{equation}
N=\sum_{k=0}^{h}r^k=\frac{1-r^{h+1}}{1-r}.
\end{equation}

\begin{figure}[!t]
    \centerline{\includegraphics[width=.99\linewidth]{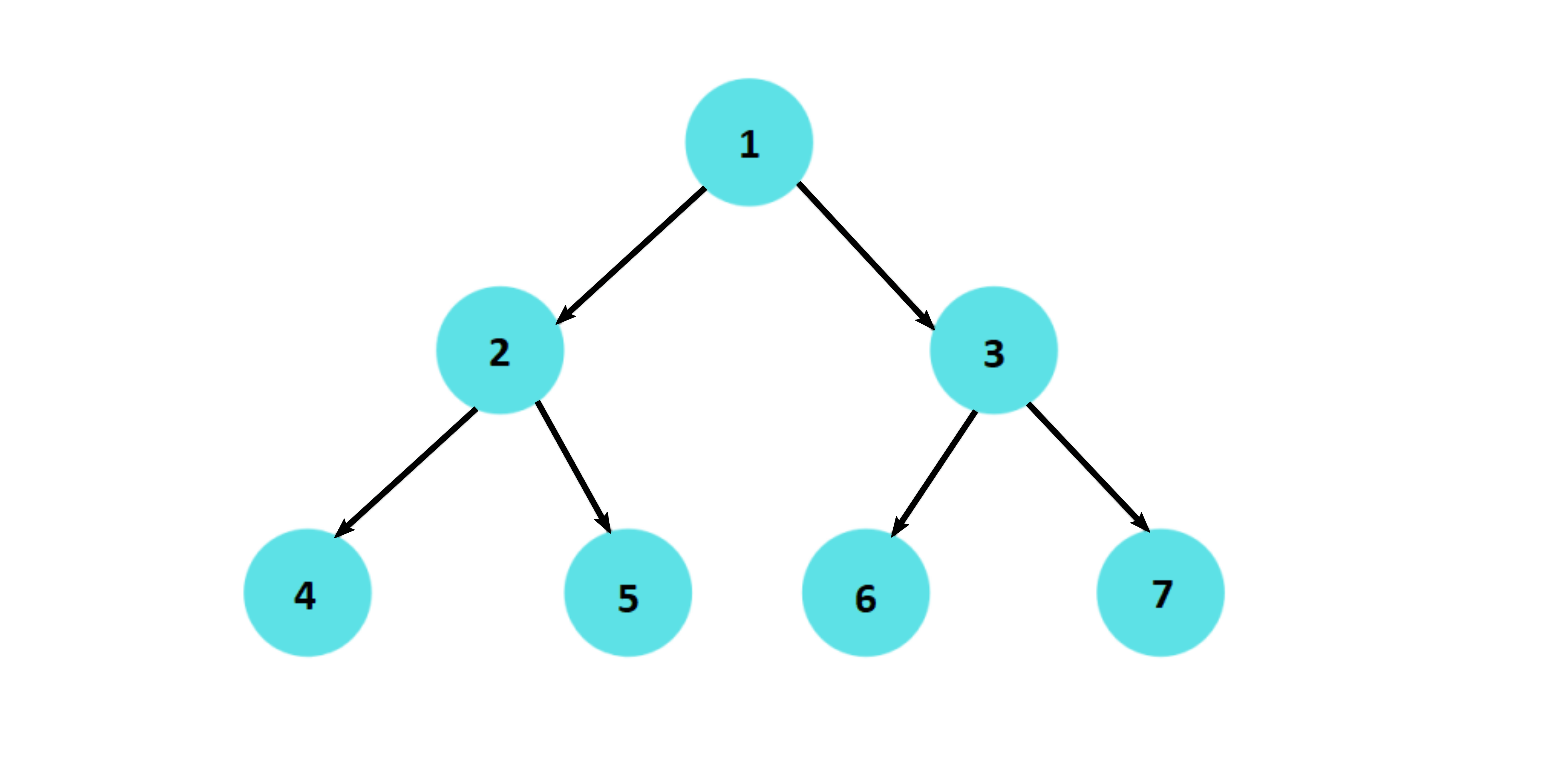}}
        \caption{Illustration of a perfectly balanced binary polytree.}
    \label{figbalanced}
    \vspace{-20pt}
\end{figure}
\begin{figure}[!t] 
  \label{ fig7} 
  \begin{minipage}[b]{0.5\linewidth}
    \centering
    \includegraphics[width=.99\linewidth]{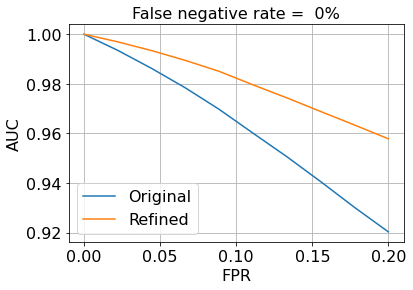} 
  %  \caption{FNR=0\%.} 
  \\ \centering (a) FNR=0\%.
    \label{fnr0}
  \end{minipage}%%
  \begin{minipage}[b]{0.5\linewidth}
    \centering
    \includegraphics[width=.99\linewidth]{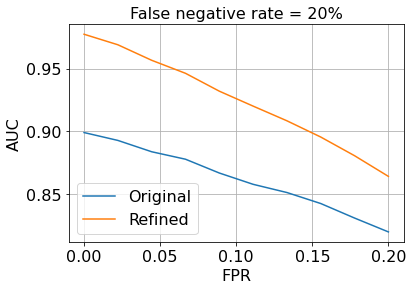} 
  %  \caption{FNR=20\%.} 
  \\ \centering (b) FNR=20\%.
    \label{fnr20}
  \end{minipage} 
  \begin{minipage}[b]{0.5\linewidth}
    \centering
    \includegraphics[width=.99\linewidth]{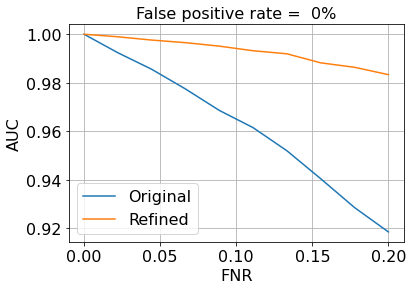} 
   % \caption{FPR=0\%.} 
   \\ \centering (c) FPR=0\%.
    \label{fpr0}
  \end{minipage}%% 
  \begin{minipage}[b]{0.5\linewidth}
    \centering
    \includegraphics[width=.99\linewidth]{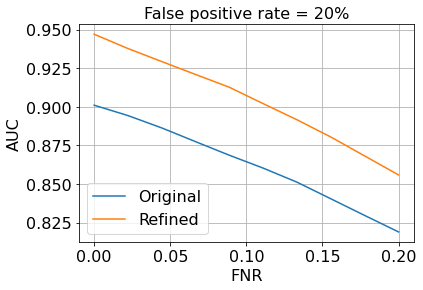} 
  %  \caption{FPR=20\%.} 
  \\ \centering (d) FPR=20\%.
    \label{fpr20}
  \end{minipage} 
  \caption{Performance for perfectly balanced binary polytree.}
  \label{balancedtreeresults}
  \vspace{-15pt}
\end{figure}

In our numerical implementations, we will focus on $(r,h)$-polytrees for which the edges are always directed downwards in the tree. An illustration of such a perfectly balanced $(2,2)$-polytree is reported in Fig. \ref{figbalanced}. Next, we describe the behavior of the anomalies scores set $\mathcal{S}$. Assuming that an anomaly detector has been selected for identifying network anomalies, we consider that at each time epoch $m$, a set of KPIs $\mathcal{R}_m\subset V$ exhibits anomalous behavior. However, given the possible unreliability of the detector in question, we suppose that there exists a non-zero FPR and FNR. In particular, for each node $i\in\mathcal{R}_m$, there's a probability FNR that the score $s_{mi}$ will be equal to $0$ instead of $1$. Additionally, for each node $i\in V\setminus\mathcal{R}_m$, there's a probability FPR that the score $s_{mi}$ will be equal to $1$ instead of $0$. It is worth noting that, in our numerical implementation, the set $\mathcal{R}_m$ is sampled uniformly at each time epoch $m$ from all the possible paths of length $h$ in $G$ and that the number of time epochs $M$ is set to $5000$. To evaluate the performance advantage that our refinement block brings to the anomaly detector, we compare the performance of the overall anomaly detection framework in both the presence and absence of our refinement block. As a performance measure, we adopt the Area Under Curve - Receiver Operating Characteristic (AUC-ROC) metric, widely used in classification problems \cite{Hand2001}. Particularly, this performance measure illustrates the diagnostic ability of a classifier system in separating anomalous KPIs behavior from a normal one. The AUC ranges from $0$ to $1$, which corresponds to a poor and excellent score respectively. With this in mind, let us first consider that the experts' graph is a $(2,6)$-polytree, and let us fix $\overline{\alpha}$ to $0.2$. Next, we investigate two distinct scenarios: 1) we fix the FNR and vary the FPR, and 2) we fix the FPR and vary the FNR. In both cases, we plot the AUC of both the original anomaly score $\mathcal{S}$ and the refined counterpart $\mathcal{Y}$. As seen in Fig. \ref{balancedtreeresults}, our proposed refinement block consistently outperforms the original score set $\mathcal{S}$ in terms of AUC, showcasing its performance advantage in improving the KPIs anomaly detection framework. It is worth noting that our proposed algorithm execution time at each time epoch was $200$ ms for $N=127$, showcasing its running-time efficiency. 
\subsection{Effect of $r$ and $h$}
In this section, the goal is to investigate the effect of the experts' graph $G$ structure on the performance advantage of our refinement block. To that end, we start our analysis by studying the effect of the polytree parameter $r$ on the performance. Particularly, we fix the FPR and FNR to $10\%$ and we compare the AUC score of both the original and refined anomaly scores in function of $r$. The results are reported in Table \ref{AUCm}.
\begin{center}
\begin{tabular}{|c|c|c|c|}
 \hline
 $r$ & $h$ & $\text{AUC}_{\text{Original}}$ & $\text{AUC}_{\text{Refined}}$ \\
  \hline
 $3$  & $4$  & $0.9$ & $0.929$ \\
 $4$  & $4$  & $0.9$ & $0.918$\\
   $3$  & $5$  & $0.9$ & $0.927$ \\
   $4$  & $5$  & $0.9$ & $0.918$ \\
 \hline
\end{tabular}
 \captionof{table}{AUC comparison in function of $r$.}
 \label{AUCm}
\end{center}
As can be seen, although our refined scores always outperform the original ones, the performance advantage slightly decreases with the number of children of internal nodes $r$. The reason behind this trend is that the portion of leaf nodes in the overall graph $G$ gets high as $r$ increases. Particularly, let us define the density of leaf nodes $d_{\text{leaf}}$ as the ratio of the number of leaf nodes over the total number of nodes. In an $(r,h)$-polytree, we have
\begin{equation}
    d_{\text{leaf}}=\frac{|\mathcal{L}|}{N}=\frac{r^h(1-r)}{1-r^{h+1}}=(1-\frac{1}{r})(\frac{r^h-1}{r^h}).
    \label{leafnodesnumber}
\end{equation}
From eq. (\ref{leafnodesnumber}), we can see that the density of leaf nodes increases with $r\geq1$. However, the issue with leaf nodes is that they are inherently sensitive to false positive data. Specifically, a false positive at a leaf node cannot be remedied by our block given that the constraints reported in eq. (\ref{constraintneighborhood}) are restricted to the nodes $i\in V: \mathcal{N}_i\neq\emptyset$. Accordingly, the refinement block is particularly sensitive to such issues. Given that the leaf nodes' density increases with $r$, the part of these issues increases in the overall test. Consequently, the overall AUC performance improvement with respect to the original score decreases. 

\begin{figure}[!t] 
  \label{ fig7} 
  \begin{minipage}[b]{0.5\linewidth}
    \centering
    \includegraphics[width=.99\linewidth]{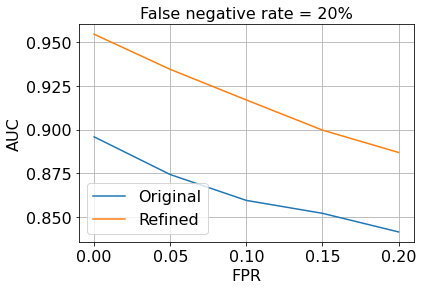} 
    \\ \centering (a) FNR=20\%.
    %\caption{FNR=20\%.} 
    \label{fnr20-true}
  \end{minipage}%%
  \begin{minipage}[b]{0.5\linewidth}
    \centering
        \includegraphics[width=.99\linewidth]{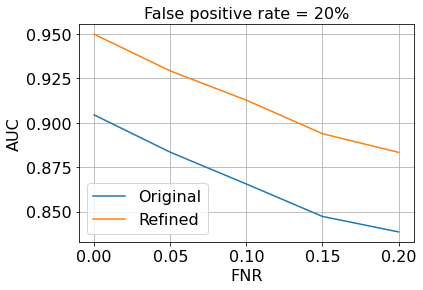} 
    %\caption{FPR=20\%.} 
    \\ \centering (b) FPR=20\%.
    \label{fpr02-true}
  \end{minipage} 
  \caption{Performance for our developed experts' graph.}
  \label{expertgraphsresultfigure}
  \vspace{-15pt}
\end{figure}

Next, we investigate the effect of the polytree height $h$ on the performance. Similarly, we fix the FPR and FNR to $10\%$ and we compare in Table \ref{AUCh} the AUC score of both the original and refined anomaly scores in function of $h$. 
\begin{center}
\begin{tabular}{|c|c|c|c|}
 \hline
 $r$ & $h$ & $\text{AUC}_{\text{Original}}$ & $\text{AUC}_{\text{Refined}}$ \\
  \hline
 $2$  & $4$  & $0.9$ & $0.937$ \\
 $2$  & $6$  & $0.9$ & $0.941$\\
   $2$  & $8$  & $0.9$ & $0.944$ \\
 \hline
\end{tabular}
 \captionof{table}{AUC comparison in function of $h$.}
 \label{AUCh}
\end{center}
As can be observed, our refined scores always outperform the original ones and the performance advantage slightly increases with $h$. To understand this trend, we need to consider the following trade-off. When $h$ increases, for a fixed number of FNs and FPs, there exists a higher chance for our block to remedy these issues since the number of True Positives (TPs) and True Negatives (TNs) also increases. However, when $h$ increases, the number of FPs and FNs also increases, worsening the performance of our refinement block. This trade-off plays a big role in determining the overall AUC performance of our refined scores in function of $h$. However, as seen in our implementations, the scale tips to the former, and the AUC performance improves with $h$. Note that the issue with leaf nodes is less prominent when $h$ increases since the density $d_{\text{leaf}}$ is upper bounded by $(1-\frac{1}{r})$ for high $h$. To conclude, the above two results suggest that the structure of the experts' graph $G$ plays a vital role in determining the degree of performance advantage the refinement block will bring. Therefore, a natural future research direction of our work is to theoretically derive performance guarantees of our proposed refinement block for a family of experts' graphs such as polytrees and polyforests.
\subsection{Practical Implementations}
\label{expertsgraphimplementation}
Thus far, we have evaluated our refinement block on directed acyclic graphs $G=(V,E)$ belonging to the polytree family. However, in practice, the graph representing the causal relationships between the various network KPIs may not belong to this family. To demonstrate the practical benefits of our approach, our team has created an expert graph that captures all network KPIs and their causal relationships. The graph includes KPIs such as CQI, Modulation and Coding Scheme (MCS), and DL/UL Block Error Rate (BLER), among others. Due to the large size of the set $V$ and $E$, we omit the full details of the graph. Nonetheless, we implement our refinement block using the same settings described in Section \ref{performancecomparisonsection} and report our results in Fig \ref{expertgraphsresultfigure}. As the graph shows, our refinement block consistently improves the performance of the anomaly detection framework in terms of AUC-ROC, further emphasizing its practical benefits.
\section{Conclusion}
\label{concsection}
In this paper, we have introduced an optimization framework for refining anomaly scores by leveraging side information in the form of an experts' graph representing the different causality relationships between the data features. We have provided a theoretical analysis of the smoothness properties of the ensuing objective function, and have proven the convergence of the proposed optimization algorithm. We have also showcased the performance gain that our proposed refinement block brings in terms of AUC-ROC compared to the original anomaly scores. With this in mind, our future research directions consist of theoretically deriving performance guarantees for our algorithm for a family of experts' graphs (e.g., polytrees).
\bibliographystyle{IEEEtran}
\bibliography{References}
\appendices
\section{Proof of Lemma \ref{lemmabeta}}
\label{prooflemmabeta}

To obtain our desired results, our goal is to show the Lipschitz continuity of the gradient $\nabla\overline{f}(\boldsymbol{z}_m,\boldsymbol{\varepsilon}_m)$ for $(\boldsymbol{z}_m,\boldsymbol{\varepsilon}_m)\in\mathbb{R}^N\times\mathbb{R}^N$. To that end, we first recall several properties of Lipschitz functions:
\begin{itemize}
    \item The sum of Lipschitz functions is also Lipschitz.
    \item The product of two bounded Lipschitz functions is also Lipschitz.
    \item The composition of Lipschitz functions is Lipschitz.
\end{itemize}
With the above properties in mind, we start by investigating the function reported in eq. (\ref{firsteq}). Given that $\overline{\alpha}>0$, it is straightforward that the first term is Lipschitz continuous. As for the second term, we first note that the smooth max function introduced in eq. (\ref{softmax}) is a bounded Lipschitz continuous \cite{2017arXiv170400805G}. Accordingly, the second term is made of a sum and product of Lipschitz continuous functions. Therefore, it is Lipschitz continuous. It is worth mentioning that introducing the cubic term in eq. (\ref{finalunconstrainedobjective})  allowed for the term $[y_{mi}-\frac{\sum_{j\in\mathcal{N}_i}y_{mj}\exp(cy_{mj})}{\sum_{j\in\mathcal{N}_i}\exp(cy_{mj})}]_{+}$ to remain in the derivative as seen in eq. (\ref{firsteq}). This enabled the Lipschitz continuity property to hold for this term, hence showcasing its importance for preserving the smoothness of $\overline{f}(\cdot,\cdot)$. As for eq. (\ref{secondeq}), the Lipschitz continuity is also straightforward given that $\overline{\alpha}>0$. Next, for the remaining terms reported in eq. (\ref{thirdeq}) and eq. (\ref{fourtheq}), we rewrite them as follows
\begin{equation}
   \frac{\partial y_{mk}}{\partial z_{mk}}=\frac{\exp(-z_{mk})}{\big(1+\exp(-z_{mk})\big)^2}=\sigma(z_{mk})(1-\sigma(z_{mk})),
\end{equation}
\begin{equation}
    \frac{\partial \alpha_{mk}}{\partial \varepsilon_{mk}}=(1-\overline{\alpha})\sigma(\varepsilon_{mk})(1-\sigma(\varepsilon_{mk})).
\end{equation}
Given that $\sigma(\cdot)$ is a bounded Lipschitz function, we can conclude that the same can be concluded about $\frac{\partial y_{mk}}{\partial z_{mk}}$ and $\frac{\partial \alpha_{mk}}{\partial \varepsilon_{mk}}$. Lastly, we note that the functions reported in eq. (\ref{firsteq})-(\ref{fourtheq}) are all bounded. Consequently, by noting the derivatives chain rule, and the fact that the composition of Lipschitz functions is also Lipchitz, we can conclude that, overall, $\nabla\overline{f}(\boldsymbol{z}_m,\boldsymbol{\varepsilon}_m)$ is a Lipschitz function. Given this property, we recall that from the mean value theorem, and for $(\boldsymbol{z}_m,\boldsymbol{\varepsilon}_m)\in \mathbb{R}^N\times\mathbb{R}^N$ and $(\boldsymbol{z}'_m,\boldsymbol{\varepsilon}'_m)\in\mathbb{R}^N\times\mathbb{R}^N$, we have
\begin{equation}
    ||\nabla\overline{f}(\boldsymbol{x})-\nabla\overline{f}(\boldsymbol{y})||\leq \beta ||\boldsymbol{x}-\boldsymbol{y}||,
\end{equation}
where $\boldsymbol{x}=[\boldsymbol{z}_m,\boldsymbol{\varepsilon}_m]$, $\boldsymbol{y}=[\boldsymbol{z}'_m,\boldsymbol{\varepsilon}'_m]$, and $\beta$ denotes the maximum eigenvalue, in magnitude, of the Hessian matrix $\boldsymbol{H}_{\overline{f}}$ over the domain $\mathbb{R}^N\times\mathbb{R}^N$. Given that the gradient $\nabla\overline{f}(\cdot,\cdot)$ is Lipschitz, we can conclude that there exists a constant $L$ such that $\beta\leq L$. This concludes our proof.

% WE NEED BOUNDNESS HERE SINCE WE ARE GONNA USE CHAIN RULE. Now, given that $\sigma(\cdot)$ is a Lipschitz continuous function, and given that the composition of Lipschitz functions is also Lipchitz, we can conclude that, overall, the functions reported in eq. (\ref{firsteq}) and (\ref{secondeq}) are Lipchitz continuous in $(\boldsymbol{z}_m,\boldsymbol{\varepsilon}_m)$.
% Therefore, overall, we obtain that $\nabla\overline{f}(\boldsymbol{z}_m,\boldsymbol{\varepsilon}_m)$ is a Lipschitz function. 

\end{document}